\documentclass[conference]{IEEEtran}

\usepackage{amsfonts}
\usepackage{amsmath,amscd}
\usepackage{amssymb}
\usepackage{graphicx}%
\usepackage{color}
\usepackage{tikz}	
\usetikzlibrary{backgrounds,fit,decorations.pathreplacing}  

\newtheorem{theorem}{Theorem}

\newenvironment{proof}[1][Proof]{\noindent\textbf{#1.} }{\hfill \rule{0.6em}{0.6em}\\}

 \DeclareMathOperator{\tr}{Tr}

\newcommand{\proj}[1]{| #1 \rangle\!\langle #1 |}

\newcommand{\be}{{\mathbf e}}

\def\cA{{\cal A}}        

\def\cD{{\cal D}}

        \def\cK{{\cal K}}
\def\cL{{\cal L}}

\def\cM{{\cal M}}

\def\cU{{\cal U}}
        
\def\cX{{\cal X}}
\def\cY{{\cal Y}}        \def\cZ{{\cal Z}}

\def\0{{\mathbf{0}}}
\def\1{{\mathbf{1}}}
\def\2{{\mathbf{2}}}
\def\3{{\mathbf{3}}}
\def\4{{\mathbf{4}}}
\def\5{{\mathbf{5}}}
\def\6{{\mathbf{6}}}

\def\7{{\mathbf{7}}}
\def\8{{\mathbf{8}}}
\def\9{{\mathbf{9}}}

\def\be{\begin{equation}}
\def\ee{\end{equation}}
\def\bea{\begin{eqnarray}}
\def\eea{\end{eqnarray}}

\DeclareMathOperator{\Tr}{Tr}

\begin{document}

\title{Source Compression with a Quantum Helper}

\author{
\IEEEauthorblockN{Min-Hsiu Hsieh}
\IEEEauthorblockA{
University of Technology, Sydney\\
Email: Min-Hsiu.Hsieh@uts.edu.au}
\and
\IEEEauthorblockN{Shun Watanabe}
\IEEEauthorblockA{
University of Tokushima \&  University of Maryland \\
Email: shun-wata@is.tokushima-u.ac.jp }}

\maketitle

\begin{abstract}
We study classical source coding with quantum side-information where the quantum side-information
is observed by a helper and sent to the decoder via a classical channel. We derive a single-letter characterization of the achievable rate region for this problem. The direct part of our result is proved via the measurement compression theory by Winter. Our result reveals that a helper's scheme that separately conducts a measurement and a compression is suboptimal, and the measurement compression is fundamentally needed to achieve the optimal rate region.
\end{abstract}


\section{Introduction}

Source coding normally refers to the information processing task that aims to reduce the redundancy exhibited when multiple copies of the same source are used.  In establishing information theory, Shannon demonstrated a fundamental result that source coding can be done in a \emph{lossless} fashion;  namely, the recovered source will be an exact replica of the original one when the number of copies of the source goes to infinity \cite{Shannon:1948wk}. If representing the source by a random variable $X$ with output space $\cX$ and distribution $p_X$, lossless source coding is possible if and only if the compression rate $R$ is above its Shannon entropy:
\begin{equation}
R\geq H(X),
\end{equation}
where $H(X):=\sum_{x\in\cX} -p_X(x) \log p_X(x)$.

Redundancy can also exist in the scenario in which multiple copies of the source are shared by two or more parties that are far apart.  Compression in this particular setting is called \emph{distributed} source coding, which has been proven to be extremely important in the internet era.  The goal is to minimise the information sent by each party so that the decoder can still recover the source faithfully.  Shannon's lossless source coding theorem can still be applied individually to each party. However, it is discovered that a better source coding strategy exists if the sources between different parties are correlated. Denote $X$ and $Y$ the sources held by the two distant parties, where the joint distribution is $P_{XY}$ and the output spaces are $\cX$ and $\cY$, respectively. 
Slepian and Wolf showed that lossless distributed source coding is possible when the compression rates $R_1$ and $R_2$ for the two parties satisfy \cite{Slepian:1973wj}:
\begin{align}
R_1 &\geq  H(X|Y),\\
R_2  &\geq H(Y|X),\\
R_1+R_2  &\geq H(XY),
\end{align}
where $H(X|Y)$ is the conditional Shannon entropy.  This theorem is now called the classical Slepian-Wolf theorem \cite{Slepian:1973wj}. In particular, when source $Y$ is directly observed at the decoder, the problem is sometimes called source coding with (full) side-information. 

Another commonly encountered scenario in a communication network is that a centralised server exists and its role is to coordinate all the information processing tasks, including the task of source coding, between the nodes in this network. Obviously, the role of the server is simply as a helper and it is not critical to reproduce the exact information communicated by the server.  This slightly different scenario results in a completely different characterisation of the rate region, as observed by Wyner \cite{Wyner:1975iv}
and Ahlswede-K\"orner \cite{Ahlswede:1975ea}. Consider that the receiver wants to recover the source $X$ with the assistance of the server (that we will call a helper from now on) holding $Y$, where the distribution is $P_{XY}$. Wyner showed that the optimal rate region for lossless source coding of $X$ with a classical helper $Y$ is the set of rate pairs $(R_1,R_2)$ such that 
\begin{align}
R_1 & \geq  H(X|U), \label{eq:classical-region-R1} \\
R_2 & \geq  I(U;Y), \label{eq:classical-region-R2} 
\end{align}
for some conditional distribution $p_{U|Y}(u|y)$, and $I(U;Y)$ is the classical mutual information between random variables $U$ and $Y$. When there is no constraint  on $R_2$ (i.e. $R_2$ can be as large as it can be), this problem reduces to source coding with (full) side-information.

The problem of source coding,  when replacing classical sources with quantum sources, appears to be highly nontrivial in the first place\footnote{The quantum source coding result takes a much longer time to develop if one considers that quantum theory began to evolve in the mid-1920s.}. The first quantum source coding theorem was established by Schumacher \cite{Schumacher:1995dg, Jozsa:1994ea}.  A quantum source $\rho_A$ can be losslessly compressed and decompressed if and only if the rate $R$ is above its von Neumann entropy\footnote{The subscript $A$ is a label to which the quantum system $\rho_A$ belongs.}:
\begin{equation}
R\geq H(A)_\rho, 
\end{equation}
where $H(A)_\rho:= -\tr \rho_A \log \rho_A$. 

Schumacher's quantum source coding theorem bears a close resemblance to its classical counterpart. One will naturally expect that the same will hold true for the distributed source coding problem in the quantum regime. Consider that Alice, who has the quantum system $A$ of an entangled source $\rho_{AB}$, would like to merge her state to the distant party Bob.  Then, the rate $R$ at which quantum states with density matrix $\rho_A$ can be communicated to a party with quantum side information $\rho_B$ is given by the conditional von Neumann entropy $H(A|B)_\rho$, a simple observation followed from the classical Slepian-Wolf theorem. While this naive conclusion turns out to be correct, this result has a much deeper and profound impact in the theory of quantum information as it marks a clear departure between classical and quantum information theory.   It is rather perplexing that the rate $R$ is quantified by the conditional entropy $H(A|B)_\rho$, which can be negative. This major piece of the puzzle was resolved with the interpretation that if the rate is negative, the state can be merged, and in addition, the two parties will gain  $|H(A|B)_\rho|$ amount of entanglement for later quantum communication \cite{Horodecki:2005fv, Horodecki:2006hl, Dupuis:2014jz}. The distributed quantum source coding problem was later fully solved \cite{Abeyesinghe:2009ej,  Datta:2011vc} where the trade-off rate region between the quantum communication and the entanglement resource is derived.  The result is now called the fully quantum Slepian-Wolf theorem (FQSW).

Source coding with hybrid classical-quantum systems $\rho_{XB}$ with $X$ representing a classical system and $B$ a quantum state is also considered in quantum information theory, and our result falls into this category.  
In \cite{Devetak:2003kd}, Devetak and Winter considered classical source coding with quantum side information at the decoder, and showed that the optimal rate $R_1$ is given by $H(X|B)_\rho$. This result can be regarded as a classical-quantum version
of the source coding with (full) side-information.


In this work, we consider classical source coding with a quantum helper, a problem that was completely overlooked before.  In our problem, the quantum side-information is observed by the helper, and  the decoder will only have a classical description from the quantum helper.  Although our problem can be regarded as a classical-quantum version of the classical helper problem studied in \cite{Wyner:1975iv, Ahlswede:1975ea}, in contrast to its classical counterpart,  our problem does not reduce to source coding with quantum side-information studied in \cite{Devetak:2003kd} even if there is no constraint on rate $R_2$.  However, when the ensemble that constitutes the quantum side-information is commutative, our problem reduces to the classical helper problem.

We completely characterize the rate region of the quantum helper problem. In fact, the formulae describing the rate region (cf.~Theorem \ref{Theorem:rate-region}) resembles its classical counterpart 
(cf.~\eqref{eq:classical-region-R1} and \eqref{eq:classical-region-R2}).  However, the proof technique is very different due to the quantum nature of the helper.   In particular, we use the measurement compression theory by Winter \cite{Winter:2004uk} in the direct coding theorem. One of interesting consequences of our result is that a helper's scheme that separately conducts a measurement and
a compression is suboptimal; measurement compression is fundamentally needed to achieve the optimal rate region.

There are a huge amount of work devoted to both classical and quantum lossy source coding \cite{Shannon:1959tf, Berger71, Devetak:2002it, Datta:2013ur, Wilde:2013hp, Datta:2013jk}. However, we will restrict ourselves to only noiseless source coding in this work.  

\emph{Notations.} In this paper, we will use capital letters $X,Y,Z,U$ etc.~to denote classical random variables, and lower cases $x,y,z,u$  to denote their realisations. We use $\cX,\cY,\cZ,\cU$ to denote the sample spaces. We denote $x^n=x_1x_2\cdots x_n$. 

A quantum state is a positive semi-definite matrix with trace equal to one. We will use $\rho$ or $\sigma$ to denote a quantum system in this paper. In case we need to specify which party the quantum state belongs to, we will use a subscript description $\rho_A$, meaning that the quantum system is held by A(lice). Letting $\{\proj{x}\}_{x\in\cX}$ be a set of orthonormal basis vectors, a classical-quantum state $\rho_{XB}$ is written as
\begin{align*}
\rho_{XB}=\sum_{x} p_{X}(x) \proj{x} \otimes \rho_{x},
\end{align*}
so that $n$ copies of it is
\begin{align*}
\rho_{XB}^{\otimes n}=\sum_{x^n} p_{X}^{(n)}(x^n) \proj{x^n} \otimes \rho_{x^n},
\end{align*}
where we denote $\rho_{x^n}:=\rho_{x_1}\otimes\cdots\otimes\rho_{x_n}$ for the sequence $x^n$. A positive-operator valued measure (POVM), $\Lambda=\{\Lambda_y\}$, is a quantum measurement whose elements are non-negative self-adjoint operators on a Hilbert space so that $\sum_{y\in\cY}\Lambda_y = I$.

This paper is organised as follows. In Sec~\ref{secII}, we formally define the problem of source coding with a quantum helper, and present the main result as well as its proof. We conclude in Sec~\ref{secIII} with open questions.

\section{Classical Source Compression with a Quantum Helper}\label{secII}
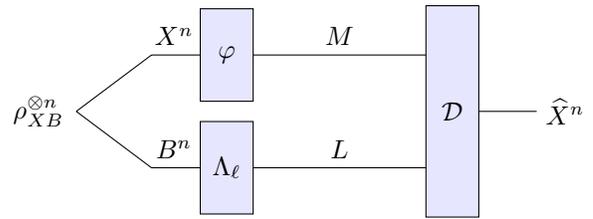
\begin{figure}
\centerline{
    \begin{tikzpicture}[scale=1][very thick]
    \fontsize{10pt}{1} 
    \tikzstyle{halfnode} = [draw,fill=white,shape= underline,minimum size=1.0em]
    \tikzstyle{checknode} = [draw,fill=blue!10,shape= rectangle,minimum height=3.5em, minimum width=2em]
    \tikzstyle{checknode2} = [draw,fill=blue!10,shape= rectangle,minimum height=8em, minimum width=2em]
    \tikzstyle{variablenode} = [draw,fill=white, shape=circle,minimum size=0.8em]
    \node (e1) at (-2.5,1) {$\rho_{XB}^{\otimes n}$} ;
    \node (p3) at (4.5,1) {$\widehat{X}^n$} ;
    \node (w2) at (1.5,2) {$M$} ;
    \node (l2) at (1.5,0.5) {$L$} ;
    \node (s2) at (-0.7,2) {$X^n$} ;
    \node (e2) at (-0.7,0.5) {$B^n$} ;
     \draw  (-2,1)-- (-1,1.75) (-2,1)-- (-1,0.25) (-1,0.25) -- (0,0.25) (-1,1.75) -- (0,1.75);
     \draw (3.3,1) --++ (p3);
     \draw (0.3,0.25) --(3,0.25) (0.3,1.75)--(3,1.75);
    \node[checknode] (cn1) at (0,1.75) {${\varphi}$};
    \node[checknode2] (cn2) at (3,1) {${\cal D}$};
    \node[checknode] (cn3) at (0,0.25) {${\Lambda}_{\ell}$};  
   \end{tikzpicture}
 }

  \caption{Source Compression with a Quantum Helper.
  }\label{fig:QSCQH}
\end{figure}

As shown in Figure~\ref{fig:QSCQH}, the protocol for classical source coding with a quantum helper involves two senders, Alice and Bob, and one receiver, Charlie. Initially Alice and Bob hold $n$ copies of a classical-quantum state $\rho_{XB}$.
In this case, Alice holds classical random variables $X^n$ while Bob (being a helper) holds a quantum state $\rho_{X^n}$ that is correlated with Alice's message. The goal is for the decoder Charlie to faithfully recover Alice's message when assisted by the quantum helper Bob.

We now proceed to formally define the coding procedure. We define an $(n,\epsilon)$ code for classical source compression with a quantum helper to consist of the following:
\begin{itemize}
\item Alice's encoding operation $\varphi: \cX^n \to \cM$, where $\cM:=\{1,2,\cdots, |\cM|\}$ and $|\cM|=2^{nR_1}$;
\item Bob's POVM $\Lambda=\{\Lambda_\ell\}: B^n \to \cL$, where $\cL:=\{1,2,\cdots,|\cL|\}$ and $|\cL|=2^{nR_2}$;
\item  Charlie's decoding operation $\cD:\cM\times\cL\to \widehat{\cX}^n$
\end{itemize}
so that the error probability satisfies
\begin{align}
\Pr\{X^n\neq \widehat{X}^n\}\leq \epsilon.
\end{align}

A rate pair $(R_1,R_2)$ is said to be \emph{achievable} if for any $\epsilon,\delta>0$ and all sufficiently large $n$, there exists an $(n,\epsilon)$ code with rates $R_1+\delta$ and $R_2+\delta$. The rate region is then defined as the collection of all achievable rate pairs. Our main result is the following theorem. 

\begin{theorem} \label{Theorem:rate-region}
Given is a classical-quantum source $\rho_{XB}$. The optimal rate region for lossless source coding of $X$ with a quantum helper $B$ is the set of rate pairs $(R_1,R_2)$ such that 
\begin{align}
R_1 &\geq  H(X|U) \\
R_2 &\geq  I(U;B)_\sigma. 
\end{align}
The state $\sigma_{UB}(\Lambda)$ resulting from Bob's application of the POVM $\Lambda=\{\Lambda_u\}_{u\in\cU}$ is 
\begin{align}
\sigma_{UB}(\Lambda) = \sum_{u\in\cU} p_U(u) \proj{u}\otimes \rho_u
\end{align}
where
\begin{align}
p_U(u) &= \Tr(\rho_B\Lambda_u) \\
\rho_u &= \frac{1}{p_U(u)} [\sqrt{\rho_B} \Lambda_u \sqrt{\rho_B}]^* \\
\rho_B&= \sum_x p_X(x)\rho_x.
\end{align}
where $*$ denotes complex conjugation in the standard basis. Furthermore, we can restrict 
the size of POVM as $|\cU| \le d_B^2$, where $d_B$ is the dimension of Bob's system. 
\end{theorem}

\begin{figure}[t]
\centering{
\begin{minipage}{.4\textwidth}
\includegraphics[width=\textwidth]{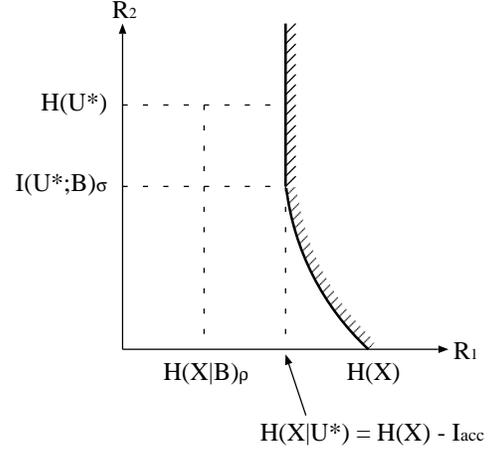}
\caption{A typical shape of the rate region in Theorem \ref{Theorem:rate-region}.}
\label{Fig:region}
\end{minipage}
}
\end{figure}

A typical shape of the rate region in Theorem \ref{Theorem:rate-region} is 
described in Fig.~\ref{Fig:region}. When there is no constraint on $R_2$, rate $R_1$ can be decreased as small as
\begin{align}
H(X|U^*) &:= \min_{\Lambda} H(X|U) \\
&= H(X) - \max_{\Lambda} I(X; U) \\
&= H(X) - I_{\mathrm{acc}},
\end{align}
where $I_{\mathrm{acc}}$ is the accessible information for the ensemble $\{ (p_X(x), \rho_x) \}_{x\in{\cal X}}$.
Unless the ensemble commutes \cite{Hayden:2004},
the minimum rate $H(X|U^*)$ is larger than the rate $H(X|B)_\rho$, which is the optimal rate
in the source coding with quantum side-information \cite{Devetak:2003kd}. 
To achieve $R_1 = H(X|U^*)$, it suffices to have $R_2 \ge I(U^*;B)_\sigma$, which is smaller than
$H(U^*)$ in general. This means that the following separation scheme is suboptimal:
first conduct a measurement to get $U^*$ and then compress $U^*$.
For more detail, see the direct coding proof.

\begin{proof}[Converse]

Let $\varphi: {\cal X}^n \to {\cal M}$ be Alice's encoder, and let
$\{ \Lambda_{\ell} \}_{\ell \in {\cal L}}$ be Bob's measurement. 
Alice sends $M = \varphi(X^n)$ to the decoder, and Bob sends the measurement outcome $L$ to the decoder. The Fano's inequality states that $H(X^n|M,L)\leq n\epsilon_n$ for some $\epsilon_n\to 0$ as $n\to \infty$.

First, we have the following bound:
\begin{align}
\log |{\cal M}| 
&\ge H(M) \\
&\ge H(M|L) \\
&\ge H(X^n|L) - H(X^n|M,L) \\
&\stackrel{(a)}\ge H(X^n|L) - n \epsilon_n \\
&\stackrel{(b)}\ge \sum_{t=1}^n H(X_t|X_{<t},L) - n \epsilon_n \\
& \stackrel{(c)}= \sum_{t=1}^n H(X_t|U_t) - n \epsilon_n \\
& \stackrel{(d)}= n H (X_J|U_J, J) - n \epsilon_n,
\end{align}
where $(a)$ follows from Fano's inequality: $H(X^n|M,L)\leq n \epsilon_n$ for some $\epsilon_n\to 0$ as $n \to \infty$; in (b), we use chain rule and denote $X_{<t} := (X_1,\ldots,X_{t-1})$; in $(c)$, we denote $U_t:=(X_{<t},L)$; in $(d)$, we introduce a time-sharing random variable $J$ that is uniformly distributed in the set $\{1,2,\cdots n\}$.

Next, we have
\begin{align}
\log |{\cal L}|
&\ge H(L) \\
&\ge I(L ; B^n) \\
&= \sum_{t=1}^n I(L; B_t|B_{<t}) \\
&= \sum_{t=1}^n I(L, B_{<t}; B_t) \\
&\stackrel{(a)}{=} \sum_{t=1}^n I(L, B_{<t}, X_{<t}; B_t) \\
&\ge \sum_{t=1}^n I(L, X_{<t}; B_t) \\
& = \sum_{t=1}^n I(U_t; B_t) \label{eq_L00}
\end{align}
where (a) follows from 
\begin{align}
I(X_{<t} ; B_t|L,B_{<t}) 
&\le I(X_{<t}; B_t,B_{>t}|L,B_{<t}) \\
&= H(X_{<t}|L,B_{<t}) - H(X_{<t}|L,B^n) \\
&\le H(X_{<t}|B_{<t}) - H(X_{<t}|L,B^n) \\
&= H(X_{<t}|B_{<t}) - H(X_{<t}|B^n) \\
&= H(X_{<t}|B_{<t}) - H(X_{<t}|B_{<t}) \\
&= 0.
\end{align}

Following from Eq.~(\ref{eq_L00}), we can again introduce a time-sharing random variable $J$ that is uniformly distributed in the set $\{1,2,\cdots, n\}$, 
\begin{align}
\sum_{t=1}^n I(U_t; B_t) & = n \sum_{t=1}^n  I(U_t; B_t |J=t) \\
&= n I(U_J; B_J|J) \\
&= n I(U_J J;B_J)
\end{align}
where the last equality follows because $I(J;B_J)=0$. To get single-letter formula, define $X=X_J$, $B=B_J$, and $U=(U_J,J)$ and let $n\to \infty$:
\begin{align}
R_1 &= \frac{1}{n} \log |\cM| \geq H(X|U) \\
R_2 &=\frac{1}{n} \log|\cL| \geq I(U;B).
\end{align}

Here, we note that the distribution of $U_t = (L,X_{<t})$ can be written as
\begin{multline}
p_{X_{<t} L}(x_{<t},\ell) =  \left(\prod_{i < t} p_X(x_i)\right) \times \\
  \Tr\left[ \left\{ \left( \bigotimes_{i < t} \rho_{x_i} \right) \otimes \rho_{B_t}
  \otimes \left( \bigotimes_{i > t} \rho_{B_i} \right)\right\}  \Lambda_\ell  \right].
\end{multline}
Thus, we can get $U_t$ as a measurement outcome of $B_t$ by first generating $X_{<t}$, then
by appending $\bigotimes_{i < t} \rho_{x_i}$ and $\bigotimes_{i > t} \rho_{B_i}$ to ancillae systems, 
and finally by conducting the measurement $\{ \Lambda_\ell \}_{\ell \in {\cal L}}$.

Finally, the bound on $|{\cal U}|$ can be proved via Carath\'odory's theorem (cf.~\cite[Appendix C]{Devetak:2005ea}).


\end{proof}

\begin{proof}[Direct Coding Theorem]
Fix a POVM measurement $\Lambda=\{\Lambda_u\}_{u\in\cU}$. It induces a conditional probability $p_{U|X}(u|x)= \Tr [\Lambda_u\rho_x]$, and joint probability distribution 
\begin{equation}\label{eq_Pau}
P_{XU}(x,u)=p_X(x)p_{U|X}(u|x).
\end{equation} The crucial observation is the application of Winter's measurement compression theory \cite{Winter:2004uk}. 

\begin{theorem}[Measurement compression theorem \cite{Winter:2004uk, Wilde:2012iq}] \label{thm:meas-comp}Let $\rho_A$ be a source state and $\Lambda$ a POVM\ to simulate on this state. A protocol for a
faithful simulation of the POVM\ is achievable with classical communication rate $R$ and common randomness rate $S$ if and only if the following set of inequalities hold%
\begin{align}
R   \geq I\left(  X;R\right) ,~~
R+S   \geq H\left(  X\right)  ,
\end{align}
where the entropies are with respect to a state of the following form:%
\begin{equation}
\sum_{x}\left\vert x\right\rangle \left\langle x\right\vert ^{X}%
\otimes \text{Tr}_{A}\left\{  \left(  I^{R}\otimes\Lambda_{x}%
^{A}\right)  \phi^{RA}\right\}  ,\label{eq:IC-state}%
\end{equation}
and $\phi^{RA}$ is some purification of the state $\rho_A$.
\end{theorem}

Let $K$ be a random variable on ${\cal K}$, which describes the common randomness shared between
Alice and Bob. Let 
$\{ \widetilde{\Lambda}_{u^n}^{(k)} \}_{u^n \in {\cal U}^n}$ be collection of POVMs. Let 
\begin{align}
Q_{X\widetilde{U}}^n(x^n,u^n) := P_X^{(n)}(x^n) \sum_{k \in {\cal K}} \frac{1}{|{\cal K}|} {\rm{Tr}}[ \rho_{x^n} \widetilde{\Lambda}_{u^n}^{(k)}],
\end{align}
where $P_X^{(n)}(x^n):=P_X(x_1)\times\cdots\times P_X(x_n)$.
The faithful simulation of $n$ copies of POVM $\Lambda:=\{ \Lambda_u \}_{u\in {\cal U}}$, i.e.~$\Lambda^{\otimes n}$, implies that for any $\epsilon>0$,  there exists $n$ sufficiently large, such that there exist POVMs $\{\widetilde{\Lambda}^{(k)}\}$, where $\widetilde{\Lambda}^{(k)}:=\{\widetilde{\Lambda}_{u^n}^{(k)} \}_{u^n \in {\cal U}^n}$, with  
\begin{align} \label{eq:faithful-simulation}
\frac{1}{2} \| P_{XU}^{(n)} - Q_{X\widetilde{U}}^{n} \|_1 \le \epsilon. 
\end{align}

\noindent\textbf{Coding Strategy:} 

Alice and Bob shared  $n$ copies of the state $\rho_{XB}$, and assume that Bob performs measurement $\Lambda^{\otimes n}: B^{\otimes n}\to \cU^n$ on his quantum system whose outcome is sent to the decoder to assist decoding Alice's message. Bob's measurement  on each copy of $\rho_{XB}$ will induce the probability distribution $P_{XU}$ according to (\ref{eq_Pau}). Apparently, if Bob sends the full measurement outcomes to Charlie (say $nH(U)$ bits), then Charlie can successfully decode $X^n$ simply from Slepian-Wolf Theorem. The next strategy is to make use of classical result since after Bob's measurement, Alice and Bob become fully classical with joint distribution $P_{XU}$. Therefore, the minimum rate for Bob is $I(V;U)$ (w.r.t.~some conditional distribution $p_{V|U}(v|u)$). However, there is a non-trivial quantum coding strategy. Detail follows.

\emph{Bob's coding.} Instead of the measurement $\Lambda$ performed on Bob's system $\rho_B$ and coding w.r.t.~the classical channel $p_{V|U}(v|u)$, the decoder Charlie can directly simulate the measurement outcome $U$ using Winter's  measurement compression theorem  \cite{Winter:2004uk, Wilde:2012iq}. Denote Bob's classical communication rate $R_2=\frac{1}{n}\max_{k\in\cK}|\widetilde\Lambda^{(k)}|$. Then Theorem~\ref{thm:meas-comp} promises that by sending $R_2\geq I(U;B)$ from Bob to the decoder Charlie, Charlie will have a local copy $\widetilde{U}^{ n}$ and the distribution between Alice and Charlie $Q^n_{X\widetilde{U}}$ will satisfy (\ref{eq:faithful-simulation}). 

\emph{Alice's coding.} Now Alice's strategy is very simple since Charlie has had $\widetilde{U}^n$. She just uses the Slepian-Wolf coding strategy as if she starts with the distribution $P_{XU}$ with Charlie.
In fact, it is well known (cf.~\cite{Cover})
that there exists an encoder $\varphi:\cX^n \to \cM$ and a decoder 
$\cD:\cM \times \cU^n \to \cX^n$ such that $|\cM| = 2^{n(H(X|U)+\delta)}$ and 
\begin{align}
P^{(n)}_{XU}({\cal A}^c) \le \epsilon
\end{align}
for sufficiently large $n$, where 
\begin{align}
{\cal A} := \{ (x^n,u^n) \in \cX^n \times \cU^n: \cD(\varphi(x^n),u^n) = x^n \}
\end{align}
is the set of correctably decodable pairs.

Now, suppose that Alice and Bob use the same code for the simulated distribution $Q_{X\widetilde{U}}^n$.
Then, by the definition of the variational distance and \eqref{eq:faithful-simulation}, we have
\begin{align}
 Q^n_{X\tilde{U}}({\cal A}^c) \le P^{(n)}_{XU}({\cal A}^c) + \epsilon.
\end{align}
Thus, if we can find a good code for $P_{XU}^{(n)}$, we can also use that code for $Q^n_{X\widetilde{U}}$ for sufficiently large $n$.

\emph{Derandomization.} The standard derandomization technique works here. Since the distribution $Q^{n}_{X\widetilde{U}} = \frac{1}{|\cK|} \sum_{k\in\cK}Q^n_{X\widetilde{U}|k}$, and
\begin{align}
\sum_k \frac{1}{|{\cal K}|} Q^n_{X\widetilde{U}|K=k}({\cal A}^c) = 
 Q^n_{X\tilde{U}}({\cal A}^c) \le P_{XU}({\cal A}^c) + \epsilon.
\end{align}
Thus, there exists one $k\in\cK$ so that $Q^n_{X\widetilde{U}|k}(\cA_n^c)$ is small. 

\end{proof}

\section{Conclusion and Discussion}\label{secIII}

We considered the problem of compression of a classical source with a quantum helper. We completely characterised its rate region and showed that the capacity formula does not require regularisation, which is not common in the quantum setting.  While the expressions for the rate region are similar to the classical result in \cite{Wyner:1975iv, Ahlswede:1975ea, ElGamal:2011ty}, it requires vey different proof technique. To prove the achievability, we employed a powerful theorem, measurement compression theorem \cite{Winter:2004uk}, that can decompose quantum measurement. A similar approach was recently applied to derive a non-asymptotic bound on  the classical helper problem \cite{Watanabe:2013ea}.

This work brings more questions than answered. As we have pointed out, source coding with a helper was never considered in the quantum regime before ours. Our work can be served as the first step to the more general (fully quantum) setting; namely, quantum source coding with a quantum helper. Currently, it is completely unknown how to quantify the distinction between side information and a quantum helper. We believe that resolving this question will sharpen our understanding of a quantum source. 


\section*{Acknowledgements}
MH is supported by an ARC Future Fellowship under Grant FT140100574. 
SW is supported in part by JSPS Postdoctoral Fellowships for Research Abroad.


\begin{thebibliography}{99}


\bibitem{Shannon:1948wk} C. E. Shannon, ``A mathematical theory of communication,'' Bell Syst. Tech. J., vol. 27, pp. 379--423, 623--656, 1948.

\bibitem{Slepian:1973wj} D. Slepian and J. K. Wolf, ``Noiseless coding of correlated information sources,'' IEEE Trans. Inform. Theory, vol. 19, no. 4, pp. 471--480, 1973.

\bibitem{Wyner:1975iv} A. D. Wyner, ``On source coding with side information at the decoder,'' IEEE Trans. Inform. Theory, vol. 21, no. 3, pp. 294--300, 1975.

\bibitem{Ahlswede:1975ea} R. Ahlswede and J. K\"orner, ``Source coding with side information and a converse for the degraded broadcast channel," IEEE Trans. Inform. Theory, vol. 21, no. 6, pp. 629--637, 1975.

\bibitem{Schumacher:1995dg} B. Schumacher, ``Quantum coding,'' Phys. Rev. A, vol. 51, no. 4, pp. 2738--2747, Apr. 1995.

\bibitem{Jozsa:1994ea} R. Jozsa and B. Schumacher, ``A New Proof of the Quantum Noiseless Coding Theorem,'' J. of Modern Optics, vol. 41, no. 12, pp. 2343--2349, Dec. 1994.

\bibitem{Hayden:2004} P. Hayden, R. Jozsa, D. Petz, and A. Winter, ``Structure of States Which Satisfy Strong Subadditivity of Quantum Entropy with Equality,'' Communications in Mathematical Physics, vol. 246, no. 2, pp. 359--374, Feb. 2004.

\bibitem{Horodecki:2005fv} M. Horodecki, J. Oppenheim, and A. Winter, ``Partial quantum information,'' Nature, vol. 436, no. 7051, pp. 673--676, Aug. 2005.

\bibitem{Horodecki:2006hl} M. Horodecki, J. Oppenheim, and A. Winter, ``Quantum State Merging and Negative Information,'' Communications in Mathematical Physics, vol. 269, no. 1, pp. 107--136, Oct. 2006.

\bibitem{Dupuis:2014jz} F. Dupuis, M. Berta, J. Wullschleger, and R. Renner, ``One-Shot Decoupling,'' Communications in Mathematical Physics, vol. 328, no. 1, pp. 251--284, Mar. 2014.

\bibitem{Abeyesinghe:2009ej} A. Abeyesinghe, I. Devetak, P. M. Hayden, and A. Winter, ``The mother of all protocols: restructuring quantum information's family tree,'' Proceedings of the Royal Society A: Mathematical, Physical and Engineering Sciences, vol. 465, no. 2108, pp. 2537--2563, Jun. 2009.

\bibitem{Datta:2011vc} N. Datta and M.-H. Hsieh, ``The apex of the family tree of protocols: optimal rates and resource inequalities,'' New J. Phys., vol. 13, no. 9, p. 093042, 2011.

\bibitem{Devetak:2003kd} I. Devetak and A. Winter, ``Classical data compression with quantum side information,'' Phys. Rev. A, vol. 68, no. 4, Oct. 2003.

\bibitem{Shannon:1959tf} C. E. Shannon, ``Coding theorems for a discrete source with a fidelity criterion,'' IRE Nat. Conv. Rec, vol. 4, pp.~142--163, 1959.

\bibitem{Berger71} T. Berger, \emph{Rate Distortion Theory: A Mathematical Basis for Data Compression}. Englewood Cliffs, NJ: Prentice Hall, 1971.

\bibitem{Devetak:2002it} I. Devetak and T. Berger, ``Quantum rate-distortion theory for memoryless sources,'' IEEE Trans. Inform. Theory, vol. 48, no. 6, pp. 1580--1589, 2002.

\bibitem{Datta:2013ur} N. Datta, M.-H. Hsieh, and M. M. Wilde, ``Quantum Rate Distortion, Reverse ShannonTheorems, and Source-Channel Separation,'' IEEE Trans. Inform. Theory, vol. 59, no. 1, pp. 615--629, 2013.

\bibitem{Wilde:2013hp} M. M. Wilde, N. Datta, M.-H. Hsieh, and A. Winter, ``Quantum Rate-Distortion Coding With Auxiliary Resources,'' IEEE Trans. Inform. Theory, vol. 59, no. 10, pp. 6755--6773, 2013.

\bibitem {Datta:2013jk} N. Datta, M.-H. Hsieh, M. M. Wilde, and A. Winter, ``Quantum-to-classical rate distortion coding,'' J. Math. Phys., vol. 54, no. 4, p. 042201, 2013.

\bibitem{Cover} T. M. Cover and J. A. Thomas, \emph{Elements of Information Theory }. Wiley, New York, 1991.

\bibitem{ElGamal:2011ty} A. El Gamal and Y.-H. Kim, Network information theory. Cambridge University Press, 2011.

\bibitem{Winter:2004uk} A. Winter, ``Extrinsic'' and ``Intrinsic'' Data in Quantum Measurements: Asymptotic Convex Decomposition of Positive Operator Valued Measures. Communications in Mathematical Physics, 244(1), 157--185, 2004.

\bibitem{Wilde:2012iq} M. M. Wilde, P. M. Hayden, F. Buscemi, and M.-H. Hsieh, ``The information-theoretic costs of simulating quantum measurements,'' Journal of Physics A: Mathematical and Theoretical, vol. 45, no. 45, pp. 453001, Nov. 2012.

\bibitem{Watanabe:2013ea} S. Watanabe, S. Kuzuoka, and V. Y. F. Tan, ``Non-Asymptotic and Second-Order Achievability Bounds for Source Coding With Side-Information," in Proc. 2013 IEEE International Symposium on Information Theory, pp. 3055--3059.

\bibitem{Devetak:2005ea} I. Devetak and A. Winter, ``Distillation of secret key and entanglement from quantum states," in Proc. of The Royal Society A, vol. 461, pp. 207--235, Jan. 2005.

\end{thebibliography}
\end{document}